\documentclass[useAMS]{mn2e}
\usepackage{graphicx}

\def\apj{ApJ}

\def\aap{A\&A}

\def\prd{Phys. Rev. D}

\title[Structure formation under extended gravity] {A first linear cosmological structure formation scenario under extended gravity} 

\author[X. Hernandez and M.A. Jim\'enez] {X. Hernandez and  M. A. Jim\'enez \\ 
Instituto de Astronom\'{\i}a, Universidad Nacional Aut\'{o}noma de M\'{e}xico,
  Apartado Postal 70--264 C.P. 04510 M\'exico D.F. M\'exico.}

\begin{document}

\maketitle

\begin{abstract}
The inability of primordial baryonic density fluctuations, as observed in the cosmic microwave background (CMB), to grow
into the present day astronomical structures is well established, under Newtonian and Einsteinian gravity. It is hence customary
to assume the existence of an underlying dark matter component with density fluctuations, $\Delta(M)$, having amplitudes much larger 
than what CMB observations imply for the baryons. This is in fact one of the recurrent arguments used in support of the dark
matter hypothesis. In this letter we prove that the same extended theory of gravity which has been recently shown to accurately
reproduce gravitational lensing observations, in absence of any dark matter, and which in the low velocity regime converges
to a MONDian force law, implies a sufficiently amplified self-gravity to allow purely baryonic fluctuations with amplitudes in 
accordance with CMB constraints to naturally grow into the $z=0$ astrophysical structures detected. The linear structure formation 
scenario which emerges closely resembles the standard concordance cosmology one, as abundantly calibrated over the last decade to 
match multiple observational constraints at various redshifts. However, in contrast with what occurs in the concordance cosmology,
this follows not from a critical dependence on initial conditions and the fine tuning of model parameters, but from the rapid 
convergence of highly arbitrary initial conditions onto a well defined $\Delta(M,z)$ attractor solution.
\end{abstract}


\section{Introduction} 

In Bernal et al. (2011) a relativistic extended gravity model was presented, which working under a FLRW metric was recently shown 
in Carranza et al. (2013) to be consistent with the observed expansion history of the Universe, including the recent
accelerated expansion phase. In Mendoza et al. (2013) we proved that the same relativistic extended gravity scenario, working
under an spherically symmetric, static Schwarzschild-like metric, results in a gravitational lensing framework in full
accordance with the observed phenomenology, all the above considering exclusively baryonic matter as inferred from
observations, without the need of any dark components. The relativistic extended gravity model of Bernal et al. (2011),
by construction, converges in the low velocity limit to a MONDian force law, as required to explain galactic rotation curves
e.g. Milgrom (1983), Famaey \& McGaugh (2012), observed stellar dynamics of dwarf galaxies e.g. McGaugh \& Wolf (2010), 
Hernandez et al. (2010), and the recently measured outer flattening of globular cluster dispersion velocity profiles e.g. 
Scarpa et al. (2011), Hernandez et al. (2013), in the absence of any dark matter.

Since under the standard gravity scenario, augmented by the introduction of a hypothetical dark matter component, an essentially 
constant dark matter fraction is required across astrophysical scales, it is reasonable to suspect that a model which replaces
the dark matter component by an enhanced self-gravity of the baryons, might naturally also solve the cosmological
structure formation puzzle. In this letter, working with the linearised cosmological density contrast evolution equation,
we show that indeed, replacing the Newtonian for the MONDian self-gravity expression, yields substantially faster density contrast
growth factors. For comparison, in an $a(t)=(3H_{0}t/2)^{2/3}$ universe, the growth of the density contrast changes from the
Newtonian solution of $\Delta \propto (1+z)^{-1}$, to $\Delta \propto (1+z)^{-3}$. Clearly, having 3 orders of magnitude 
in redshift since recombination, allows for growth factors of $10^{9}$, and hence purely baryonic fluctuations as
observed with $\Delta \sim 10^{-5}$ in the CMB can become amply non-linear by substantially high redshifts.

Additionally, we find that the character of the linearised cosmological density contrast evolution equation changes
qualitatively from the standard case where solutions are highly sensitive to initial conditions, to an equation
having a strong attractor solution. This last point replaces the need for delicately crafted initial conditions, to
a situation where it is the self-gravity of baryonic perturbations alone what essentially fixes the structure formation scenario.

This resulting structure formation scenario is highly reminiscent of what appears under the standard concordance cosmology, 
with a bottom up growth of astrophysical structures, but without the need of specifying a detailed primordial fluctuation spectrum, 
or of calibrating bias, anti-bias, feedback parameters, etc. The modified baryonic Jeans mass at $z_{CMB}$ is of 
$4\times 10^{5} M_{\odot}$, mass-scales which become non-linear by $z \approx 19$, which hence defines the corresponding start of 
reionization redshifts.

\section{Evolution of small density perturbations in the expanding universe}

We are interested in the growth of gravitational instabilities in the non-relativistic regime within an expanding 
universe. We shall follow the well known procedure established for the case of standard gravity e.g Longair (2008), and modify 
only the self-gravity term to use the corresponding MONDian expression. First, we write the fluid dynamical 
equations including a self-gravity term: the equation of conservation of mass, the Euler equation, and the equation for the 
self-gravitational potential generated by a density field, $\rho(\textbf{r})$:
 
\begin{equation}
\frac{d\rho}{dt}=-\rho\nabla\cdot\textbf{v}
\end{equation}

\begin{equation}
 \frac{d\textbf{v}}{dt}=-\frac{1}{\rho}\nabla p- \nabla\phi
\end{equation}

\begin{equation}
\phi =\phi(\rho, \textbf{r})
\end{equation}

In the Newtonian case, eq.(3) is the standard Poisson equation with an explicit dependence only on $\rho$. Note that this equations 
are written in Lagrangian form.

Considering a homogeneous expanding background upon which a small perturbation evolves, $\textbf{v}=\textbf{v}_{0}+\delta\textbf{v} $, 
$\rho=\rho_{0}+\delta\rho$, $\phi=\phi_{0}+\delta\phi$ and $p=p_{0}+\delta p$, we can write equations (1) and (2) keeping only terms
to first order in the perturbation to yield:

\begin{equation}
\frac{d}{dt}\frac{\delta\rho}{\rho_{0}}=\frac{d\Delta}{dt}=-\nabla \cdot\delta\textbf{v}
\end{equation}

\begin{equation}
 \frac{d(\delta\textbf{v})}{dt}+(\delta\textbf{v}\cdot\nabla)\delta\textbf{v}=\frac{-1}{\rho_{0}}\nabla \delta p -\nabla\delta\phi
\end{equation}

\noindent where we use the comoving quantities

\begin{equation}
\textbf{x}=a(t)\textbf{r},
\end{equation}

\begin{equation}
 \textbf{v}=\frac{\delta x}{\delta t }= \frac{da}{dt}\textbf{r}+a(t)\frac{\textbf{r}}{dt},
\end{equation}

\noindent with $\textbf{v}_{0}=da/dt$ identified as the Hubble expansion term and $\delta\textbf{v}$ the perturbation on the 
Hubble flow, $a(t)(d\textbf{r}/dt)$. From equation (7) the perturbed velocity field, $a(t)\textbf{u}$, now results as:  

\begin{equation}
 \frac{d\textbf{u}}{dt}+2\left(\frac{1}{a}\frac{da}{dt}\right)\textbf{u}=\frac{-1}{\rho_{0}a^{2}}\nabla 
\delta p-\frac{1}{a^{2}}\nabla_{c}^{2}(\delta\phi).
\end{equation}

Considering adiabatic perturbations to replace $\delta p$ in the above equation for $c_{s}^{2}\delta\rho$,
and taking the comoving divergence of this same equation, to eliminate $\textbf{u}$ using the time derivative of equation (4) 
gives:

\begin{equation}
 \frac{d^{2}\Delta}{dt^{2}}+2\left(\frac{1}{a}\frac{da}{dt}\right)\frac{d\Delta}{dt}=\frac{c_{s}^{2}}{\rho_{0}a^{2}}\nabla_{c}^2\delta\rho
 + \nabla^{2}\delta\phi,
\end{equation}

\noindent where we have introduced the density contrast as $\Delta=\delta\rho/\rho_{0}$. In analogy with the standard result,
we begin by considering the large scale regime where the pressure term in eq.(9) can be neglected, yielding:

\begin{equation}
 \frac{d^{2}\Delta}{dt^{2}}+2\left(\frac{1}{a}\frac{da}{dt}\right)\frac{d\Delta}{dt} = \nabla^{2}\delta\phi.
\end{equation}

The quantity $\nabla^{2}\delta\phi$ depends on the theory of gravity one assumes. In the Newtonian case, 
$\nabla^{2}\delta\phi= 4\pi G \delta\rho$, but if the potential is the MONDian one introduced by Mendoza et al. (2011) 
we should write:

\begin{equation}
 \nabla\delta\phi=\frac{\sqrt{a_{0}G\delta m}}{r}.
\end{equation}

Notice that since we are working in the linear regime where the density contrast is small, we can safely assume the 
accelerations below $a_0=1.2\times10^{-8}cm/s^{2}$ limit of the extended gravity force law. For a top hat density fluctuation we can
write $ \delta m(r)=\frac{4\pi}{3} r^{3}\delta\rho$, equation (11) yields for within the fluctuation

\begin{equation}
 \nabla\delta\phi=\left(\frac{4 \pi}{3} a_{0} G r \delta\rho  \right)^{1/2}.
\end{equation}

Now we take the divergence of the gradient of this potential perturbation to obtain the Laplacian of the MONDian potential as,

\begin{equation}
 \nabla^{2}\delta\phi=\nabla\cdot\nabla\delta\phi=\left( \frac{4\pi}{3} a_{0}G\delta\rho \right)^{1/2}
\frac{1}{r^{2}}\frac{\partial}{\partial r} r^{5/2}, 
\end{equation}

\noindent giving:

\begin{equation}
 \nabla^{2}\delta\phi=\left( \frac{25 \pi}{3} \frac{a_{0}G\delta\rho}{r} \right)^{1/2}.
\end{equation}

Evaluating this last expression at the edge of the density fluctuation, we write $r= \left(\frac{3}{4\pi} \delta m/\delta\rho 
\right)^{1/3}$ where $\delta m$ is now the total fluctuation mass, to eliminate $r$ from the Laplacian of the MONDian potential, 
which yields:

\begin{equation}
 \nabla^{2}\delta\phi=\frac{5}{2} \left(\frac{4\pi}{3}\right)^{2/3} \frac{(Ga_{0})^{1/2} \rho_{0}^{2/3}}{(\delta m)^{1/6}}\Delta^{2/3}
\end{equation}

We can now study the evolution of over-densities in the linear regime in an extended gravity scenario, by substituting the result
of equation (15) into (10):

\begin{equation}
\frac{d^{2}\Delta}{dt^{2}}+2\left(\frac{1}{a}\frac{da}{dt}\right)\frac{d\Delta}{dt} = \frac{5}{2}\left(\frac{4\pi}{3}\right)^{2/3} 
\frac{(Ga_{0})^{1/2} \rho_{0}^{2/3}}{(\delta m)^{1/6}}\Delta^{2/3},
\end{equation}

\noindent which is the main result of this section. Particular solutions to the above equation and comparisons to the standard 
Newtonian results appear in the following section.

\section{Solutions for particular \lowercase {$a(t)$} cases}

We begin by examining the evolution of density fluctuations evolving within a flat universe described by:

\begin{equation}
 a(t)=\left(\frac{3H_{0}}{2}t\right)^{2/3}.
\end{equation}

\noindent This idealised case will serve merely as a test where solutions are analytical, and comparison to well known
standard results can be clearly explored. In the Newtonian case equation (10) becomes:

\begin{equation}
 \frac{d^{2}\Delta}{dt^{2}}+\left(\frac{4}{3t}\right)\frac{d\Delta}{dt}= \frac{2}{3t^{2}}\Delta,
\end{equation}

\noindent with a growing mode solution $\Delta\propto t^{2/3}\propto a = (1+z)^{-1}$.
Alternatively, when we work in the modified gravity scenario for a Universe that follows  
$a=(3H_{0}t/2)^{2/3}$ dynamics, equation (16) describing the evolution of density perturbations becomes:

\begin{equation}
\frac{d^{2}\Delta}{dt^{2}}+\left(\frac{4}{3t}\right)\frac{d\Delta}{dt}= 0.124\frac{a_{0}^{1/2}}{(G\delta m)^{1/6}}
\frac{\Delta^{2/3}}{t^{4/3}}.
\end{equation}

\begin{figure}
\includegraphics[height=7.0cm,width=8.5cm]{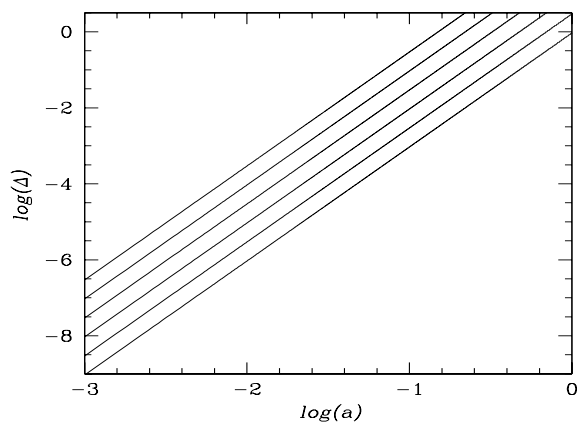}
\caption{Growth of density contrast in an $a(t)=(3H_{0}t/2)^{2/3}$ universe for the power law convergent solution of eq.(19).
Fluctuation masses of $4\times(10^{5},10^{6},10^{7},10^{8},10^{9}$ and $10^{10}) M_{\odot}$ appear in descending order.}
\end{figure}

In this last equation we have considered $H_{0}=70 km/s Mpc^{-1}$ and $\rho_{0}=0.05(3H_{0}/8\pi G)(1+z)^{3}$, the contribution 
of baryonic 
matter only. It is important to note that in this section, particular $a(t)$ scale factor evolution models are considered merely 
as convenient parametrisations of the expansion history of the Universe, as calibrated through a large number of empirical 
observations across a range of redshifts. Thus, we are not assuming a standard GR theory behind any of the $a(t)$ models tested, 
it has been shown that modified gravity theories of the $f(R)$ type can self-consistently account for the expansion histories 
obtained under GR models with parameters as calibrated to match cosmological inferences, e.g. Nojiri \& Odintsov (2011), Capozziello 
\& De Laurentis (2011), or Carranza et al. (2013) for the particular metric extended gravity theory which converges to the MONDian 
force law used here.

This time, for equation (19) there exist a unique power law solution,
$\Delta=c_{1}t^{2}$ where $c_{1}=1.898\times10^{-5}\left(a_{0}^{3}/G\delta m\right)^{1/2}$. To write $\Delta$ as a 
function of the scale factor we use $t^{2}=4a^{3}/9H_{0}$ to obtain:

\begin{equation}
\Delta=\left(\frac{M_{c}}{\delta m}\right)^{1/2}a^{3}=\left(\frac{M_{c}}{\delta m}\right)^{1/2}(1+z)^{-3},
\end{equation}

\noindent where we have introduced 

\begin{equation}
 M_{c}=7.12\times 10^{-11} \frac{a_{0}^{3}}{GH_{0}^{4}}= 3.488\times10^{10}M_{\odot}.
\end{equation}

By comparing eq.(20) to the equivalent solution in the Newtonian case which appears following eq.(18), we see that the
$(1+z)^{-1}$ scaling has been replaced by a $(1+z)^{-3}$ one. This shows that growth factors of 9 orders of magnitude, rather than
the 3 orders of the Newtonian case, will result for the interval from  $z_{CMB}\approx 1000$ to today. Thus, purely baryonic
density fluctuations with amplitudes as observed in the CMB, of order $\Delta \approx 10^{-5}$, will have ample time to naturally
grow under their own self-gravity alone into the non-linear regime, by substantially high redshifts. Therefore, the requirement
under the Newtonian approach of a hypothetical underlying undetected dark component with density fluctuations many orders
of magnitude larger than what the observed density component shows, is removed. Results for the evolution of the growth factor from 
eq.(20) are shown in figure (1), for fluctuation masses of $4\times(10^{5},10^{6},10^{7},10^{8},10^{9}$ and $10^{10}) M_{\odot}$, 
appearing in descending order. The lower fluctuation mass limit of  $4\times 10^{5} M_{\odot}$ was chosen as the baryonic
MONDian Jeans mass at recombination of $\sigma^{4} /G a_{0}$ with $\sigma$ the sound speed of 3000 K hydrogen gas, e.g. 
Mendoza et al. (2011).

\begin{figure}
\includegraphics[height=7.0cm,width=8.5cm]{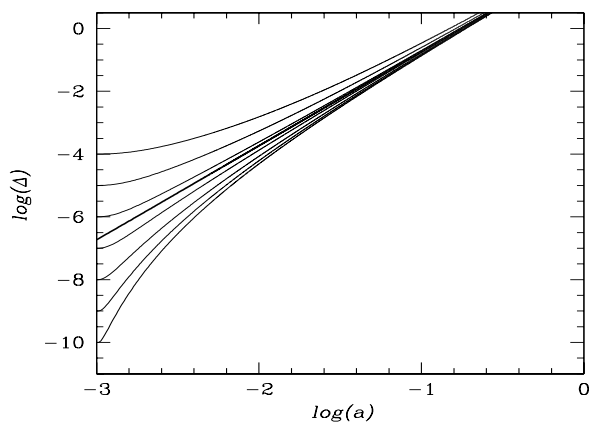}
\caption{Numerical solutions to eq.(19) for a fluctuation mass of $10^{6} M_{\odot}$, for a range of
$\Delta(z_{CMB})$ initial conditions extending over 6 orders of magnitude. Notice the strong convergence to the power law
solution of eq.(20), thick curve.}
\end{figure}

Notice also that in this case, the power law solution has a unique normalisation, as happens e.g.
when one solves for a power law solution to the hydrostatic equilibrium of a Newtonian isothermal self-gravitating gas, the
singular isothermal solution which results furnishes not only a definitive power law behaviour, but also a unique amplitude
fully determined by the physical parameters of the problem. Here, $G, a_{0}$ and $H_{0}$ fully define the amplitude and evolution of
the density contrast at all redshifts, once a fluctuation mass is chosen. This last point is related to the strongly attractive
character which the power law solution eq.(20) has. 

From the $\Delta=\Delta_{CMB}(1+z_{CMB})/(1+z)$ solutions of the Newtonian case, we see that taking different $\Delta_{CMB}$
initial conditions results in evolutionary tracks in $(\Delta, z)$ space which remain parallel throughout. Thus, initial conditions
$\Delta_{CMB}(M)$ are preserved during the linear evolutionary phase. 
The consequence of this feature is the delicate dependence of the standard structure formation scenario upon the initial
conditions, the details of which hence become crucial to determining the ensuing structure formation scenario.

The situation emerging from the modified MONDian force law in eq.(16) is thoroughly different; $\Delta$ in the source term in the
right hand side appears to a power smaller than 1, and hence if we take an enhanced solution having a slightly larger amplitude at 
a given reference redshift than a given reference solution, the source term will be proportionally smaller than the increase in
$\Delta$ itself, so that now, the reference solution will catch up with the enhanced variant. It is clear that the power law
solution to eq.(16) of eq.(20) will thus be a strongly attractive solution. This is shown explicitly in figure (2), where a number of
numerical solutions to eq.(19) are shown for a constant fluctuation mass of $10^{6} M_{\odot}$, for a range of initial conditions
at $a=10^{-3}$, covering 6 orders of magnitude, all with $d\Delta/dt =0$ at $a=10^{-3}$. The solid line shows the convergent solution 
for the same mass, of eq.(20), which is clearly a very strongly attractive solution. We thus see that the resulting structure formation
scenario will be highly independent of the initial conditions, and also, that initial density contrast values at $z_{CMB}$ in the galactic
region, much smaller than the $\Delta \sim 10^{-5}$ values observed for the extragalactic scales now measured, will be amply sufficient
to yield non-linear structures by high redshifts.

\begin{figure}
\includegraphics[height=7.0cm,width=8.5cm]{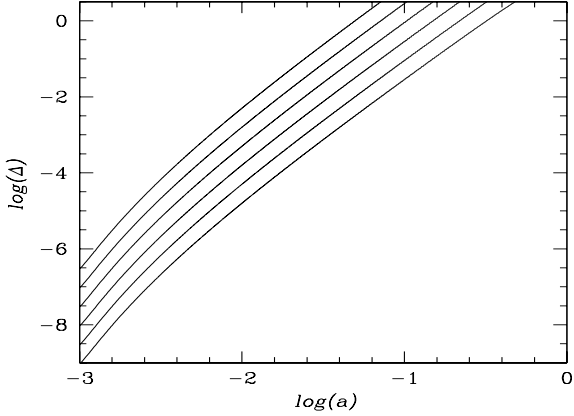}
\caption{Growth of density contrast in a universe having an $a(t)$ evolution as the concordance case, for the attractor solution to 
eq.(23). Fluctuation masses of $4\times(10^{5},10^{6},10^{7},10^{8},10^{9}$ and $10^{10}) M_{\odot}$ appear in descending order.}
\end{figure}

At this point we examine the evolution of the density contrast, but under a realistic $a(t)$ model.
The evolution of the expansion factor for a flat universe for the concordance cosmology case, as abundantly calibrated to yield
accordance with a large number of observations across a redshift range extending out to $z_{CMB}$ is:

\begin{equation}
 a(t)=\left(\frac{\Omega_{m}}{\Omega_{\Lambda}}\right)^{2/3}\left[ sinh\left(\frac{3}{2}\sqrt{\Omega_{\Lambda}}H_{0}t\right)\right]^{2/3}.
\end{equation}

As already mentioned, the above equation is taken as merely a convenient fit to the actual $a(t)$ evolution of the
Universe, which is accurately reproduced by choosing the numerical parameter values $\Omega_{m}=0.3$ and $\Omega_{\Lambda}=0.7$.
Introducing this expression in equation (16) we have: 

\begin{eqnarray}
\frac{d^{2}\Delta}{dt^{2}}+A(t)\frac{d\Delta}{dt}=B(t)\frac{\Delta^{2/3}}{dm^{1/6}}
\end{eqnarray}

\begin{figure}
\includegraphics[height=7.0cm,width=8.5cm]{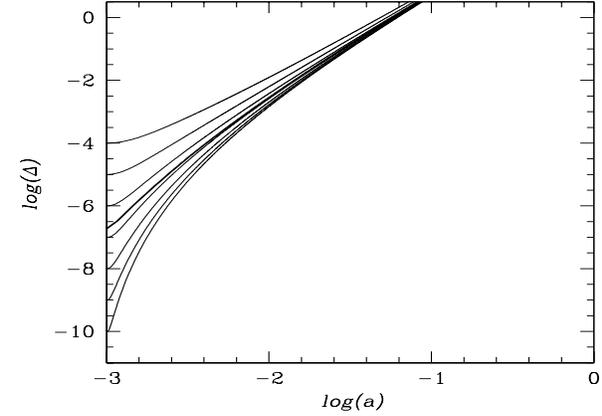}
\caption{Numerical solutions to eq.(23) for a fluctuation mass of $10^{6} M_{\odot}$, for a range of
$\Delta(z_{CMB})$ initial conditions extending over 6 orders of magnitude. Notice the strong convergence to the attractor 
solution of eq.(23), thick curve.}
\end{figure}

\noindent where:

\begin{equation}
 A(t)=\frac{2H_{0}\Omega_{\Lambda}}{tanh\left(\frac{3}{2}\sqrt{\Omega_{\Lambda}}H_{0}t\right)},
\end{equation}

\begin{eqnarray}
B(t)=\frac{c_{2}}{\left[ sinh\left(\frac{3}{2}\sqrt{\Omega_{\Lambda}}H_{0}t\right)\right]^{4/3}},
\end{eqnarray}

\noindent and $c_{2}$ is given by

\begin{equation}
 c_{2}=\frac{5}{2}\left(\frac{4\pi\Omega_{m}}{3\Omega_{\Lambda}}\right)^{2/3}\left(a_{0}G\right)^{1/2}\rho_{0}^{2/3}. 
\end{equation}

By solving eq.(23) numerically, we find again a strongly attractive solution given by taking initial conditions at $z_{CMB}$ from
the power law solution of eq.(20), which are shown in figure (3) for the same fluctuation masses appearing in figure (1).
The strongly attractive character of the solutions shown in figure (3) can again be traced to the structure of eq.(16), and is
shown explicitly in figure (4), which is analogous to figure (2). We note that for $log(a)\geqslant-2$ the growth factor evolution
shown in figure (3) can be accurately fitted by:

\begin{equation}
 \Delta= \left(\frac{M_{cr}}{\delta m}\right)^{1/2}a^{3.16}= \left(\frac{M_{cr}}{\delta m}\right)^{1/2} (1+z)^{-3.16},
\end{equation}

\noindent where this time $M_{cr}=6.5 \times 10^{13} M_{\odot}$. By comparing figure (3) to the $a=(3H_{0}t/2)^{2/3}$ case of figure (1), 
we see that for the more realistic case having an $a(t)$ evolution as that of the concordance cosmological model, the enhanced amount
of time implied by a given redshift interval now allows for substantially more growth for the density fluctuations treated. In fact,
from eq.(27), we see that the smallest primordial structures, those having the MONDian baryonic Jeans mass at $z_{CMB}$ of 
$4 \times 10^{5} M_{\odot}$, will become non-linear by a redshift of 19. This last point provides a good qualitative agreement 
with re-ionisation constraints e.g. $z=11.1 \pm 1.1$ for the redshift at which the Universe is half re-ionised of the recent Plank 
results, Planck Collaboration (2013).

Notice that our result of eq.(16) will also apply to other modified relativistic theories of gravity which in the $v<<c$
limit tend to a MONDian force law e.g. Bekenstein (2004) or Zhao \& Famaey (2010). 
We end by commenting that by merely changing the Newtonian for the MONDian self-gravity term in the density contrast evolution
equation, not only does the enhanced self-gravity results in a sufficiently amplified growth factor evolution no longer requiring
any dark matter, but also, strongly convergent solutions appear which eliminate the need for carefully tuned initial conditions.

\section{Conclusions}\label{ccl}

We have shown that if the Newtonian self-gravity term in the cosmological linear evolution fluctuation density contrast equation 
is substituted for the equivalent MONDian one, purely baryonic density perturbations with amplitudes compatible with CMB restrictions
at $z_{CMB}$ and masses ranging from $4\times 10^{5} - 4\times 10^{10} M_{\odot}$ will enter the non-linear regime 
by redshifts of between 19 and 2.2 respectively. The resulting structure formation scenario is hence highly reminiscent 
of the one appearing under the standard concordance cosmology, with a bottom up growth of cosmological structures. The
modified baryonic Jeans mass at $z_{CMB}$ is of $4\times 10^{5} M_{\odot}$ and hence the corresponding start of reionization
redshifts will be of $\approx 19$.

This eliminates the necessity of invoking a hypothetical underlying dark matter component at $z_{CMB}$ having 
density fluctuations with amplitudes several orders of magnitude above what is observed for the empirically
measured baryonic component.

A strongly convergent growth factor solution results, which also eliminates the need for an additional primordial
fluctuation generating mechanism.

\section*{acknowledgements}

Xavier Hernandez acknowledges financial assistance from UNAM DGAPA grant IN103011-3. Alejandra Jimenez acknowledges 
financial support from a CONACYT scholarship.


\end{document}